\documentstyle[epsf,epsfig]{elsart}

\begin{document}
\begin{frontmatter}

\title{Antibunching of distorted optical wave packets \\
at a beam splitter}
\author{Toralf Gruner\thanksref{1}}
\author{and Dirk--Gunnar Welsch}
\thanks[1]{Present address: Carl Zeiss,
Optical Design Dept.,
D-73446 Oberkochen, Germany}
\address{
Theoretisch-Physikalisches Institut, 
Friedrich-Schiller-Universit\"at Jena \\
D-07743 Jena, Germany }
\begin{abstract}
Interference of single-photon wave packets at a beam splitter usually 
leads to an anticorrelation of the light intensity in the two output 
ports of the beam splitter. The effect may be regarded as ``bunching'' of 
the photons at the beam splitter and has widely been interpreted as a 
result of quantum mechanical interference between the probability 
amplitudes of indistinguishable bosonic particles. Here we
show that when the wave packets are sufficiently distorted,
then the opposite behaviour is observed, i.e., simultaneous clicks
of the photodetectors in the two output ports are favoured, which
may be regarded as ``antibunching'' of the photons at the beam splitter.
\end{abstract}

\end{frontmatter}

PACS number(s): 42.25.Hz, 42.50.Ct, 42.25.Bs, 42.30.Lr 


\section{Introduction}
\label{intro}
The study of the interference behaviour of quantized light 
has not only been of fundamental interest, but has also been
important for applications, such as the use of nonclassical light
in highly sensitive interferometry. In particular,  
two-photon interference experiments have offered new
possibilities of studying fourth-order interference phenomena 
of single-photon wave packets \cite{GHOM,GM,FL,CST,SS}.
An interesting phenomenon is the anticorrelation of the 
photoelectric counts recorded in the two output channels 
of a beam splitter \cite{SSe} or an optical fibre multiport \cite{WRWZ}.
The dependence of the coincidence events 
on the optical delay between the two
interfering wave packets can be used to determine the photon tunneling 
time through an optical barrier \cite{CKS,1a,i1}.

In order to explain the anticorrelation (bunching) behaviour 
of interfering single-photon wave packets in the output channels 
of a beam splitter,
it has been argued that when two photons arrive at the beam splitter
simultaneously, then the outgoing state is obtained by superposition
of probability amplitudes, because of the indistinguishability
of the photons. Such an interpretation could imply the conclusion that 
the interference phenomenon observed is a pure quantum effect. On the
other hand it has been shown that the effect also appears for independent 
classical light fields \cite{CKPP}. 
It may therefore be expected that classical and nonclassical light fields
give rise to quantitatively rather than qualitatively
different interference structures of the coincidence-event statistics.
In this context, the question arises of 
what is the effect on the interference fringes of correlated
incoming beams and whether or not the interference fringes enable one 
to distinguish between classical and nonclassical light. 

In this paper we show that when single-photon wave packets whose
superposition gives rise to anticorrelations in the coincidence-event 
statistics in the two output ports of a beam splitter are sufficiently 
strong distorted by a multilayer dielectric slab and a  
superposition of the distorted wave packets is measured, then 
correlations can be observed in place of anticorrelations.
Such a behaviour may be regarded as a kind of ``antibunching'',  
which would be expected for fermionic particles rather than photons. 
Clearly, identifying the distorted wave packets with single photons can 
be quite questionable, as has been stressed recently \cite{i1}. 
Moreover, correlations can only be observed when the two
incoming wave packets are produced by the same source and  
the overall field is in a correlated state.  
We analyse possible interference structures and discuss criteria for 
deciding from the observed interference fringes whether the incoming 
fields are classical or not and whether they are correlated or independent
of each other.

The paper is organized as follows. In Sec.~\ref{fq} the underlying 
formalism for describing the propagation of quantized light through
multilayer dielectric structures is outlined. In Sec.~\ref{in}
a Mach--Zehnder interferometer is considered and the interference 
fringes of the output coincidence events are determined in
dependence on the distortion of the beams in the two arms.  
Finally, a summary is given in Sec.~\ref{Sum}. 

\section{Field quantization and input-output relations}
\label{fq}

To describe the interaction of quantized radiation with
(linearly responding) dielectric matter, we use the quantization
scheme developed in \cite{i2,MLBJ,i5}, which is based on the 
determination of the Green function of the classical propagation 
problem. In particular, we consider linearly polarized light propagating 
in $x$ direction through a dielectric plate consisting of $N$ $\!=$ 
$M$ $\!-$
$\!2$ layers ($M$ $\!\ge$ $\!3$). The operator of the vector potential 
can then be given by \cite{i5}
\begin{eqnarray}
\lefteqn
{\hat{A} (x)= \sum\limits_{j=1}^M \lambda_j (x) \int_0^{\infty} {\rm d} \omega 
\,\sqrt{\frac{\hbar\beta_j(\omega) }
{4 \pi c \omega \epsilon_0 \epsilon_j (\omega) {\cal A}}} 
}
\nonumber
\\ & &
\hspace{8ex}
\times
\left[ e^{i\beta_j(\omega) \omega x/c} \, \hat{a}_{j+}(x,\omega) 
+ e^{-i\beta_j(\omega) \omega x/c} \, \hat{a}_{j-}(x,\omega) 
\right] + \mbox{H.c.},
\label{2.1}
\end{eqnarray}
where $\lambda_j (x)$ $\!=$ $\!1$ if $x_{j-1}$ $\!<$ $\!x$ $\!<$ $\!x_{j}$
and $\lambda_j (x)$ $\!=$ $\!0$ otherwise, 
$ \epsilon_j (\omega)$ $\!=$ $\! n_j^2 (\omega) $ with 
$ n_j (\omega)$ $\!=$ $\!\beta_j (\omega)$ $\!+$ $\!i \gamma _j (\omega)$ 
is the complex permittivity of the $ j $th layer, and
$ \cal{A} $ is the normalization area in the $ yz $ plane
[$j$ $\!=$ $\!1$ and $j$ $\!=$ $\!M$ refer to the left- and
right-hand side free-space surroundings of the plate, respectively,
with $\epsilon_{1}(\omega)$ $\!=$ $\!\epsilon_{M}(\omega)$ $\!=$ $\!1$]. 
The quasi-mode operators $ \hat{a}_{j\pm} (x, \omega) $ inside
the plate are associated with (damped) waves propagating to the right 
and left, respectively. Outside the plate they are independent
of $ x $ and obey the familiar bosonic commutation relations for photons
in free space,
\begin{equation}
\left[ \hat{a}_{1 \pm} (\omega), \hat{a}^{\dagger}_{1 \pm} (\omega') \right]
= \delta \left( \omega- \omega'  \right) 
,
\label{2.2} 
\end{equation}
\begin{equation} 
\left[ \hat{a}_{M \pm} (\omega), \hat{a}^{\dagger}_{M \pm} (\omega') \right]
= \delta \left( \omega- \omega'  \right) 
,
\label{2.3} 
\end{equation}
\begin{equation} 
\left[ \hat{a}_{1 \pm} (\omega), \hat{a}^{\dagger}_{M \mp} (\omega') \right]
= 0,
\label{2.4}
\end{equation}
where the output operators $ \hat{a}_{1-}(\omega) $, $ \hat{a}_{M+}(\omega) $ 
are related to the input operators $ \hat{a}_{1+}(\omega) $, 
$ \hat{a}_{M-}(\omega) $ and
bosonic operators of the plate excitations, $ \hat{g}_{\pm}(\omega) $, as
\begin{equation}
\left( \begin{array}{c}
\hat{a}_{1-}(\omega) \\ \hat{a}_{M+}(\omega)
\end{array} \right)
= 
\tilde{\bf T}(\omega)
\left( \begin{array}{c}
\hat{a}_{1+}(\omega) \\ \hat{a}_{M-}(\omega)
\end{array} \right)
+
\tilde{\bf A}(\omega)
\left( \begin{array}{c}
\hat{g}_{+}(\omega) \\ \hat{g}_{-}(\omega)
\end{array} \right)\!,
\label{2.5}
\end{equation}
\begin{equation}
\left[\hat{g}_\pm(\omega),\hat{g}^\dagger_\pm(\omega')\right] 
= \delta(\omega-\omega').
\label{2.6}
\end{equation}
In Eq.~(\ref{2.5}), $ \tilde{\bf{T}} $ and $ \tilde{\bf{A}} $,
respectively, are the characteristic transformation and absorption 
matrices of the multilayer plate \cite{i5}. Note, that the elements 
of $ \tilde{\bf{T}} $ and $ \tilde{\bf{A}} $ are not 
independent of each other, but they are related to each other such 
that the consistency of the quantization scheme is ensured.

\section{Fourth-order interference}
\label{in}

Let us consider an experimental scheme of the type studied in 
\cite{CKS} (Fig.~1). Pairs of correlated (single-photon) wave packets 
produced in a parametric down conversion process are combined by a lossless
50\%:50\% beam splitter (BS) and the time-integrated output coincidences
\begin{equation}
R(s) = \xi^2 \int {\rm d}t_1 \int {\rm d}t_2 \,
\left\langle
\hat{E}^{(-)} ( t_1 )
\hat{E}^{(-)} ( t_2 )
\hat{E}^{(+)} ( t_1 )
\hat{E}^{(+)} ( t_2 )
\right\rangle
\label{3.2}
\end{equation}
are measured (the detectors PD$_1$ and PD$_2$ in Fig.~1 are assumed
to have equal efficiency $ \xi $). 
In contrast to \cite{CKS}, here each of the two wave 
packets is allowed to pass a dielectric barrier (DB~1 and DB~2 in 
Fig.~1) before impinging on the beam splitter. The position of a 
prism (P) in one arm of the interferometer is adjusted such that the 
additional path 
(2s)
compensates for the difference between the propagations 
of the two wave packets trough different barriers.

To calculate $R(s)$, Eq.~(\ref{3.2}), we first relate the fields
$\hat{E}^{\pm}(t_{1})$ and $\hat{E}^{\pm}(t_{2})$ at the two
detectors to the fields produced by the source, applying the
input--output relations (\ref{2.5}) and assuming that the 
barriers are in the ground state (zero-temperature limit). 
In a quantum-mechanical description we then perform the averaging  
$\langle\cdots\rangle$ in Eq.~(\ref{3.2}) assuming that the 
down-conversion photons are initially prepared in an entangled state
\begin{equation}
| \Psi \rangle = \int_0^{\infty} {\rm d} \Omega \,
\alpha(\Omega) \int_0^{\Omega} {\rm d}\omega \,
f(\omega) f(\Omega-\omega) \, 
\hat{a}^{\dagger}_{\rm I}(\omega) \,
\hat{a}^{\dagger}_{\rm II}(\Omega-\omega) \, | 0 \rangle,
\label{3.1}
\end{equation}
where the subscripts I and II refer to the two arms of the
interferometer (see Fig.~1), and $ \alpha(\Omega) $ and $ f(\omega) $,
respectively, denote the band-shape functions of the laser and the 
down-conversion photons, $ f(\omega) $ being centered at $ \Omega/2 $. 
After some calculation we derive
\begin{equation}
R(s) = 2 \pi^2 {\cal N}^4 \xi^2 \left[ K_0 - K_1 (s)\right], 
\label{3.3}
\end{equation}
$ {\cal N}$ $\! =$  $\!\sqrt{\hbar/(4 \pi c \epsilon {\cal A})} $, 
where
\begin{equation}
K_0 = \int_0^{\infty} {\rm d} \Omega \, \alpha^2 (\Omega)
\int_0^{\Omega} {\rm d} \omega \, \omega (\Omega\! -\! \omega)
\left|
T^{\rm (I)}_{21} (\omega) f(\omega) 
T^{\rm (II)}_{21} (\Omega\! -\! \omega) f (\Omega\! -\! \omega)
\right|^2
\label{3.4}
\end{equation}
is independent of the position of the prism, $ s $, and the 
position-dependent part $K_{1}(s)$ reads as
\begin {eqnarray}
\lefteqn{
K_1 (s) = \int_0^{\infty} {\rm d} \Omega \, \alpha^2 (\Omega)
\int_0^{\Omega} {\rm d} \omega \, \omega (\Omega - \omega)
\left| f(\omega) f (\Omega - \omega) \right|^2
}
\nonumber
\\ & &
\hspace{10ex}
\times \,
e^{2i (2\omega - \Omega) s/c}
{T^{\rm (I)}_{21}}^\ast\! (\omega)  
T^{\rm (I)}_{21} (\Omega - \omega) 
{T^{\rm (II)}_{21}}^\ast\! (\Omega - \omega)  
T^{\rm (II)}_{21} (\omega) .
\label{3.5}
\end{eqnarray}
In Eqs.~(\ref{3.4}) and (\ref{3.5}), $ T^{\rm (I/II)}_{ij} $ ($ i,j$ $\! =$
$\! 1,2 $) are the elements of the characteristic transformation matrices 
$ \tilde{\bf T}^{\rm (I/II)} $ of the two dielectric barriers 
(for analytical results for 
lossless and lossy Bragg mirrors,
respectively, see \cite{BDS} and \cite{i5}). 

Let us compare this result with 
that obtained when a classical
light source is considered which in a complete random way produces
correlated pairs of wave packets. Straightforward calculation
yields \cite{i6}
\begin{equation}
R(s)
= 4 \pi^2 \xi^2 \left[ G_0 - G_1 (s)\right] , 
\label{3.6}
\end{equation}
where
\begin{eqnarray}
\lefteqn{
\hspace*{-4ex}
G_0
= \left \langle \int_{0}^{\infty}{\rm d} \omega 
\int_{0}^{\infty}{\rm d} \omega'
\left[ 
2 \left| T^{\rm (I)}_{21} (\omega) 
E^{\rm (I)}(\omega)
\right|^2
\left|
T^{\rm (II)}_{21} (\omega') 
E^{\rm (II)}(\omega')
\right|^2
\right. \right.
}
\nonumber
\\ & &
\hspace{18ex}
+
\left|
T^{\rm (I)}_{21} (\omega) 
E^{\rm (I)}(\omega)
T^{\rm (I)}_{21} (\omega') 
E^{\rm (I)}(\omega')
\right|^2
\nonumber
\\ & & 
\hspace{20ex}
\left. \left.
+
\left|
T^{\rm (II)}_{21} (\omega) 
E^{\rm (II)}(\omega)
T^{\rm (II)}_{21} (\omega') 
E^{\rm (II)}(\omega')
\right|^2
\right] \right \rangle_{\rm cl} 
\label{3.7}
\end{eqnarray}
and
\begin{eqnarray}
\lefteqn{
\hspace*{-4ex}
G_1(s) =  2 \mbox{Re} 
\Bigg\{ \Bigg \langle 
\int_{0}^{\infty}{\rm d} \omega 
\int_{0}^{\infty}{\rm d} \omega'
e^{2 i (\omega -\omega') s/c} 
\, T^{\rm (I)}_{21} (\omega') E^{\rm (I)}(\omega') 
T^{\rm (II)}_{21} (\omega) E^{\rm (II)}(\omega) 
}
\nonumber
\\ & &
\hspace{19ex}
\times
\left[ T^{\rm (I)}_{21} (\omega) E^{\rm (I)}(\omega)
T^{\rm (II)}_{21} (\omega') E^{\rm (II)}(\omega')\right]^*  
\Bigg\rangle_{\rm cl} \, \Bigg\} .
\label{3.8}
\end{eqnarray}
Here, $ \langle\cdots\rangle_{\rm cl} $ denotes a classical averaging,
the complex-valued Fourier components of the emitted (down-conversion)
fields in the two 
arms of the interferometer, $E^{\rm (I)}(\omega)$ and $E^{\rm (II)}(\omega)$, 
being the random variables. Note that from Eqs.~(\ref{3.7}) and (\ref{3.8}) 
the relation
\begin{equation}
\frac{|G_1 (s)|}{G_0} \le \textstyle\frac{1}{2}
\label{3.9}
\end{equation}
can be derived \cite{i6}.

To perform the classical averaging, the (joint) probability distributions 
for $ E^{\rm (I)}(\omega) $ and $ E^{\rm (II)}(\omega) $ must be specified. To 
compare the quantum regime as given by the state in Eq.~(\ref{3.1}) with 
a classical regime that is as similar as possible to the quantum regime,
we assume that the probability distributions are such that
the frequencies and phases of the laser fields and the down-conversion 
fields are related to each other as $ \delta ( \Omega$ $\! -$ 
$\! \omega$ $\!-$ $\! \omega' ) $ and $ \delta ( \varphi_{\rm L}$ $\! -$ 
$\! \varphi_{\rm I}$ $\! -$ $\! \varphi_{\rm II}) $ ($ \varphi_{\rm L}$ 
is the laser phase and $ \varphi_{\rm I} $, $ \varphi_{\rm II} $ are the 
phases of the two down-conversion fields). 
In the spirit of close correspondence between classical and
quantum regime we then may assume that Eqs.~(\ref{3.7}) and (\ref{3.8}) 
can be rewritten as 
\begin{eqnarray}
\lefteqn{
\hspace*{0ex}
G_0
= 2 \int_0^{\infty} {\rm d} \omega \, \omega (\Omega\! -\! \omega)
\left|
T^{\rm (I)}_{21} (\omega) f(\omega) 
T^{\rm (II)}_{21} (\Omega\! -\! \omega) f (\Omega\! -\! \omega)
\right|^2
}
\nonumber
\\ & &
\hspace{11ex}
+ \int_0^{\infty} {\rm d} \omega \, \omega (\Omega\! -\! \omega)
\left|
T^{\rm (I)}_{21} (\omega) f(\omega) 
T^{\rm (I)}_{21} (\Omega\! -\! \omega) f (\Omega\! -\! \omega)
\right|^2 
\nonumber
\\ & &
\hspace{11ex}
+ \int_0^{\infty} {\rm d} \omega \, \omega (\Omega\! -\! \omega)
\left|
T^{\rm (II)}_{21} (\omega) f(\omega) 
T^{\rm (II)}_{21} (\Omega\! -\! \omega) f (\Omega\! -\! \omega)
\right|^2
\label{3.7a}
\end{eqnarray}
and
\begin{eqnarray}
\lefteqn{
G_1(s) =  2 \int_0^{\infty} {\rm d}  \omega \, \omega (\Omega - \omega)
\left| f(\omega) f (\Omega - \omega) \right|^2
}
\nonumber
\\ & &
\hspace{10ex}
\times
e^{2i (2\omega - \Omega) s/c}
{T^{\rm (I)}_{21}}^\ast\! (\omega)  
T^{\rm (I)}_{21} (\Omega - \omega) 
{T^{\rm (II)}_{21}}^\ast\! (\Omega - \omega) 
T^{\rm (II)}_{21} (\omega) ,
\label{3.8a}
\end{eqnarray}
where $f(\omega)$ is the same band-shape function as in Eq.~(\ref{3.1}).

The classical relations (\ref{3.7a}) and (\ref{3.8a}) and the
quantum-mechanical relations (\ref{3.4}) and (\ref{3.5}) are seen 
to mainly differ in the second and third terms in Eq.~(\ref{3.7a}). 
Physically, 
these terms 
are related to coincidence 
events caused either by the field in the arm I or the field in the 
arm II, since a classical field in one arm of the interferometer
can always be decomposed into two parts 
by means of 
the beam splitter. The
corresponding coincidence events are of course independent of the 
prism position $ s $ and hence the fringe visibility 
\begin{equation}
{\cal V} = \frac{R_{\rm max} - R_{\rm min}}{R_{\rm max} + R_{\rm min}} 
\label{3.8b}
\end{equation}
diminishes.
Clearly, in the quantum regime a single photon in one of the
arms cannot give rise to a coincidence event. 

Let us first assume that the barriers in the arms of the interferometer
are absent and the prism is translated such that the two wave packets 
do not overlap at the beam splitter, i.e., $s$ $\!\to$ $\!\infty$.
As expected, there is no $s$-dependent interference effect,
because the interference terms in Eqs.~(\ref{3.3}) and (\ref{3.6}) 
vanish, i.e., $G_{1}(s)$ $\!\to$ $\!0$ and $K_{1}(s)$ $\!\to$ $\!0$, 
as it can easily be seen from Eqs.~(\ref{3.5}) and (\ref{3.8a}), 
respectively. In the quantum regime single-photon 
wave packets impinge separately on the beam splitter, where they
are either reflected or transmitted with 50\% probability in each case.
In the classical regime the incident wave packets correspond, in a sense, 
to multiphoton states, because there is no classical analogue of 
a single-photon state. Hence, a wave packet that impinges on the
beam splitter is split into two parts of equal intensity, which
implies that coincidence events are preferably observed [see
the second and third term in Eq.~(\ref{3.7a}) -- terms that are
not observed in Eq.~(\ref{3.4})].

To study the interference behaviour, let us introduce the normalized
time-integrated coincidence events
$R^{(\rm{n})}(s)$ $\!=$ $\!R(s)/R(\infty) $ $\!=$
$[K_{0}$ $\!-$ $\!K_{1}(s)]$ $\!/$ $\!K_{0}$ [or 
$\!=$ $[G_{0}$ $\!-$ $\!G_{1}(s)]$ $\!/$ $\!G_{0}$ in the classical
regime]. From Fig.~2 it can be seen that in the quantum regime
$R^{(\rm{n})}(s)$ decreases with decreasing value of $s$ (compared
to $s$ $\!\to$ $\!\infty$) and attains a zero-value minimum. This 
behaviour can be explained from the argument of the 
indistinguishability of the two (now interfering) photons -- there 
is no ``which-way''-information available at the detectors, because 
of the wave combination at the beam splitter. The minimum is
observed for equal optical lengths of the two paths I and II 
through the interferometer (i.e., $s$ $\!=$ $\!0$ in the figure). 
It should be pointed out that a qualitatively similar behaviour is 
also observed in the classical regime, which can be explained 
using standard arguments of classical optics. The difference 
between the quantum and classical regime is that in the latter
the fringe visibility is substantially reduced. Recalling Eq.~(\ref{3.9}), 
it is easily seen that -- compared to the zero-value minimum in the
quantum regime -- in the classical case $ R^{(\rm{n})}(0)$ 
$\! =$ $\! 1/2 $ is valid.  

Let us now consider the case when a Bragg mirror is introduced in 
one arm of the interferometer (say, the second arm). In Fig.~3
the dependence on frequency of the square of the absolute value of 
the transmittance of
an assumed barrier, 
$|T_{21} (\omega)|^{2}$, and the absolute value of the band-shape 
function of an assumed time-limited wave packet, $|f(\omega)|$,
are plotted, 
$|f(\omega)|$ being centered at the middle of the band gap ($2.68$ PHz).
Note that in the Berkeley experiments \cite{CKS} the number of layers
the barrier is composed of is $ N$ $\! =$ $\! 11 $. 
Examples of $R^{(\rm n)}(s)$ for the quantum regime are shown in Fig.~4 
and discussed in detail in Ref.~\cite{i1}. Here it should be pointed 
out that when $N$ is not too large (e.g., $N$ $\!=$ $\!11$ in 
Fig.~4) $R^{(\rm n)}$ is always smaller then unity, whereas for 
sufficiently large $N$ (e.g., $N$ $\!=$ $\!41$ in Fig.~4)
it becomes also possible that $R^{(\rm n)}$ $\!>$ $\!1$. 
Clearly, such a behaviour cannot be explained from 
interference of probability amplitudes of indistinguishable 
quanta, but this is not very surprising if we recall the strong
distortion of a wave packet at such a barrier \cite{i1}. 

Since with increasing number of layers, $N$, the transmittance of the 
barrier drastically decreases, it may be difficult to observe the effect
in the scheme considered so far. However, it may easily be observed
if the laser is detuned such that the band-shape function 
$f(\omega)$ bridges over a band-gap edge of the barrier and 
sufficiently overlaps a region of (fractional) transparency. From the
interference fringes shown for the quantum regime in Fig.~5, 
in which $|f(\omega)|$ is assumed to be centered at the upper band-gap 
edge ($3.06$ PHz), it is clearly seen that the number of coincidence
events can be reduced \mbox{($R^{(\rm n)}$ $\!<$ $\!1$)}
as well as enhanced ($R^{(\rm n)}$ $\!>$ $\!1$).
Note that with decreasing $N$ the main minimum survives, whose 
position at \mbox{$s$ $\!<$ $\!0$} indicates subliminal behaviour.  

It could be thought that the effect of enhanced coincidences results
from the distortion of the wave packet at the barrier relative
to the undisturbed wave packet in the other arm of the interferometer.
That this is not the case can be seen from Fig.~6 (for the
quantum regime). In
the figure it is assumed that the two wave packets are distorted
in the same way, introducing identical Bragg mirrors in the
two arms of the interferometer (see Fig.~1).
Again, reduced and enhanced correlations are observed. In
contrast to Fig.~5 however, $R^{(\rm n)}(s)$ is now a symmetric
function of $s$, $R^{(\rm n)}(-s)$ $\!=$ $R^{(\rm n)}(s)$, because
of the symmetry of the equipment. 

For comparison, let us now consider disentangled two-photon states 
in the quantum regime or two uncorrelated classical pulses, i.e.,
light beams which are produced by independent sources. 
In this case in Eqs.~(\ref{3.3}) and (\ref{3.6}), respectively,
$ K_1 (s) $ [Eq.~(\ref{3.5})] and $ G_1 (s) $ [Eq.~(\ref{3.8})]
read as
\begin{equation}
K_1(s) 
= \left| \int_0^{\infty} {\rm d} \Omega \, \alpha (\Omega)
\int_0^{\Omega} {\rm d} \omega \, \omega
\left|
 f(\omega) \right|^2
e^{2 i \omega s/c}
{T^{\rm (I)}_{21}}^\ast\! (\omega)  
T^{\rm (II)}_{21} (\omega) \right|^2
\label{3.10}
\end{equation}
and
\begin{equation}
G_1(s) 
=  2 \left| \left \langle \int_{0}^{\infty} {\rm d} \omega \omega \, 
e^{2 i \omega s/c} 
| f(\omega) |^2 |
{T^{\rm (I)}_{21}}^\ast\! (\omega)  T^{\rm (II)}_{21} (\omega) 
\right\rangle \right|^2 .
\label{3.11}
\end{equation}
It is seen that the number of coincidence events can only be reduced,
i.e., \mbox{$R^{\rm (n)}(s)$ $\!\leq$ $\!1$}, but not enhanced,
$R^{\rm (n)}(s)$ $\!>\hspace{-2.1ex}/$ $\!1$.

Enhanced coincidences (as shown in Figs.~4 -- 6) can
therefore be regarded as being a signature of correlated beams.
However it also depends on the band-shape functions of the interfering
wave packets, whether enhanced coincidences can be observed or not.
To be more specific, let us consider the band-shape functions 
$ f^{\rm (I/II)}(\omega)$ $\! =$ 
$\! T^{\rm (I/II)}_{21} (\omega) f( \omega ) $
of the wave packets in the two input ports of the beam splitter.
The wave packets $ f^{\rm (I/II)} (\omega) $ will be said to be of type A 
if the Fourier transform   
\begin{eqnarray}
\lefteqn{
F (4 s) = \int_{- \infty}^{\infty} {\rm d} \omega \,
\left\{
e^{ 4 i s \omega/c }
\left[
\Theta\!\left( \omega + \textstyle \frac{1}{2} \Omega \right)  -
\Theta\!\left( \omega - \textstyle \frac{1}{2} \Omega \right)
\right]
\right.
}
\nonumber
\\ & &
\hspace{11ex}
\times
\left.
\left(
\textstyle\frac{1}{4} \Omega^2 - \omega^2
\right)
f^{\rm (I)} (\omega) \left( f^{\rm (II)} ( \omega)\right)^\ast  
f^{\rm (II)} (- \omega) \left( f^{\rm (I)} (- \omega)\right)^\ast 
\right\}
\label{3.12} 
\end{eqnarray}
attains non-negative values only. Otherwise, the wave packets will be 
said to be of type B. In particular, it is easily seen that when 
$|f (\omega)|$ is a Gaussian and $T^{\rm (I/II)}_{21} (\omega)$ $\!=$ 
$\!1$, then $f^{\rm (I/II)}(\omega)$ are of type A, 
whereas in the presence of a
Bragg mirror, $T^{\rm (I/II)}_{21} (\omega)$ $\!\neq$ $\!1$,
wave packets of type B can be produced.  
Recalling Eqs.~(\ref{3.5}) and (\ref{3.8a}), we find that for wave
packets of type A the effect of enhanced coincidences cannot be observed
on principle.

\section{Summary and conclusions}
\label{Sum}

We have studied the influence of wave-packet distortion at multilayer
dielectric barriers on the coincidence-event output statistics in  
two-beam interference experiments of the type described in \cite{CKS}.
Usually the observed dependence of the interference fringes on the 
optical-path difference in the interferometer indicates 
that
the wave packets prefer to arrive at 
equal output ports rather than different ones which in a sense can be
interpreted as a ``bunching'' behaviour. 
\begin{table}[h]
\begin{center}
\vspace*{2ex}
\begin{tabular}{|c|c|c|c|} \hline
 & {\bf 
Uncorrelated
} & \multicolumn{2}{|c|}{\bf 
Correlated beams
}
\\[-1ex]
 & {\bf 
beams
 } & \multicolumn{2}{|c|}{}
\\
\hline
{\bf 
Band shape 
} & Type A or B & Type A & Type B
\\ \hline
{ \raisebox{-1.5ex}{\bf Classical} } & \multicolumn{2}{|c|}
 { \raisebox{-1.5ex}{$R^{\rm (n)}(s)\leq 1$} } 
& 
 $R^{\rm (n)}(s)\leq 1$ 
\\[-1ex]
{ \raisebox{-1.5ex}{\bf beams} } & \multicolumn{2}{|c|}{} & 
 $R^{\rm (n)}(s) > 1$ 
\\[-1ex]
 & \multicolumn{2}{|c|}
{  \raisebox{1.5ex}{$ {\cal V} \le \frac{1}{3} $} } 
& 
   $ {\cal V} \le \frac{1}{2} $
\\ \hline
{ \bf Single-photon } & \multicolumn{2}{|c|}
 { \raisebox{-1.5ex}{$R^{\rm (n)}(s)\leq 1$} } 
& 
 $R^{\rm (n)}(s)\leq 1$ 
\\[-1ex]
{ \bf quantum } & \multicolumn{2}{|c|}{} & 
 $R^{\rm (n)}(s) > 1$ 
\\[-1ex]
\raisebox{-.5ex}{ \bf beams} & \multicolumn{2}{|c|}
{  \raisebox{1.5ex}{$ {\cal V} \le 1 $} } 
& 
   $ {\cal V} \le 1 $
\\ \hline
\end{tabular}
\end{center}
\vspace*{2ex}
\caption{Conditions of reduction and enhancement of the number
of coincidence events in the two output ports of the beam splitter
in the Mach--Zehnder interferometer in Fig.~1. 
}
\label{tab1} 
\end{table}
However, when
correlated wave packets are sufficiently distorted, then the
opposite effect can also be observed, i.e., different
output ports are preferably used and hence the number of
coincidence events is enhanced, in correspondence to an ``antibunching'' 
behaviour of
the wave packets at the beam splitter. The conditions under which
the number of coincidence events is reduced or enhanced are
summarized in Table~\ref{tab1}, with special emphasis on
the differences between quantum and classical light.        
 
It should be noted out that similar effects are expected
to be observed when the distortion of the wave packets is caused 
by other than multilayer devices.  
Finally, it should be pointed out that the diminished fringe visibility 
of the interference fringes in the classical regime has been derived for 
pulses produced in a random way, in close similarity to the quantum
description of the down-conversion process. If there is an appropriate 
correlation between consecutive
wave packet pairs, this visibility may  be enhanced up to 100\%
also in the classical situation \cite{i6}.

\newpage
\begin{picture}(70,500)
\put(-60,-100){\psfig{figure=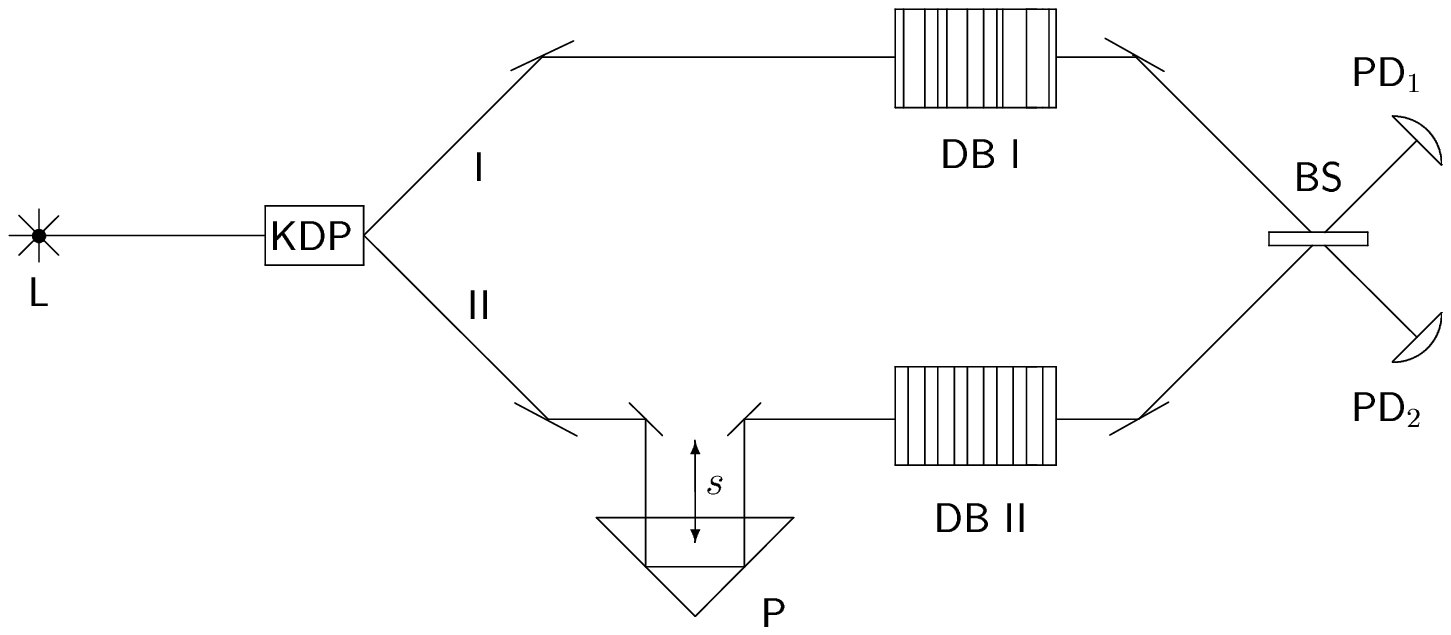,height=10in,angle=0}}
\end{picture}
\noindent
\\
Figure 1: Scheme of an extended two-photon interference experiment of the type
\protect\cite{CKS}
for determining photon traversal times through multilayer
dielectric barriers (L, laser; P, prism; DB I/II, dielectric barriers;
BS, beam splitter; PD$_1$, PD$_2$, photodetectors).

\newpage
\begin{picture}(70,500)
\put(-28,-100){\psfig{figure=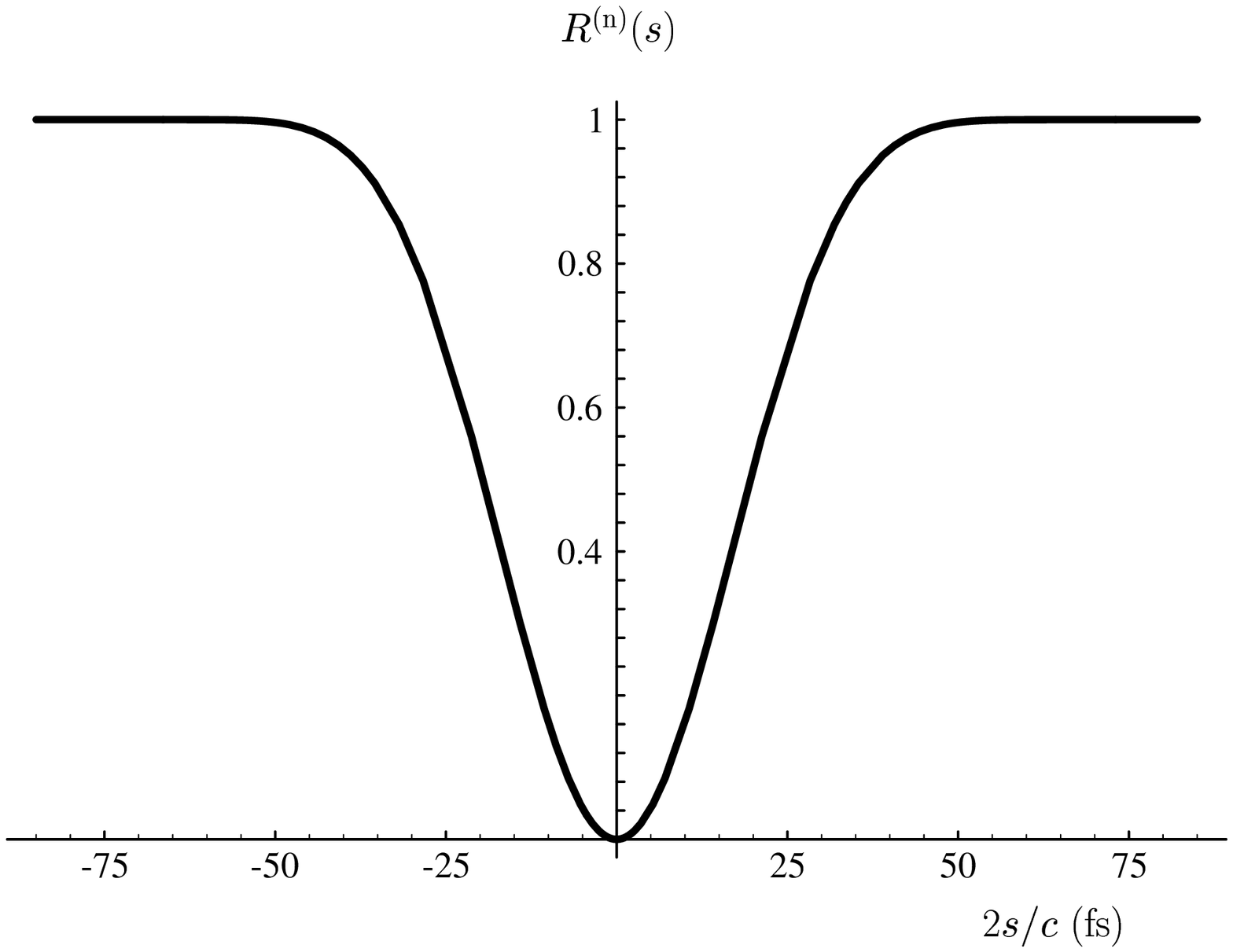,height=9in,angle=0}}
\end{picture}
\noindent
\\
Figure 2: The (normalized) coincidences $ R^{({\rm n })}(s) $ 
are shown for an experimental scheme as shown in Fig.~\protect1
in dependence on the
translation length $s$ for time-limited pulses ($2t_0$ $\!=$ $\!40$\,fs) of
photons traveling in free space.

\newpage
\begin{picture}(70,500)
\put(-28,-100){\psfig{figure=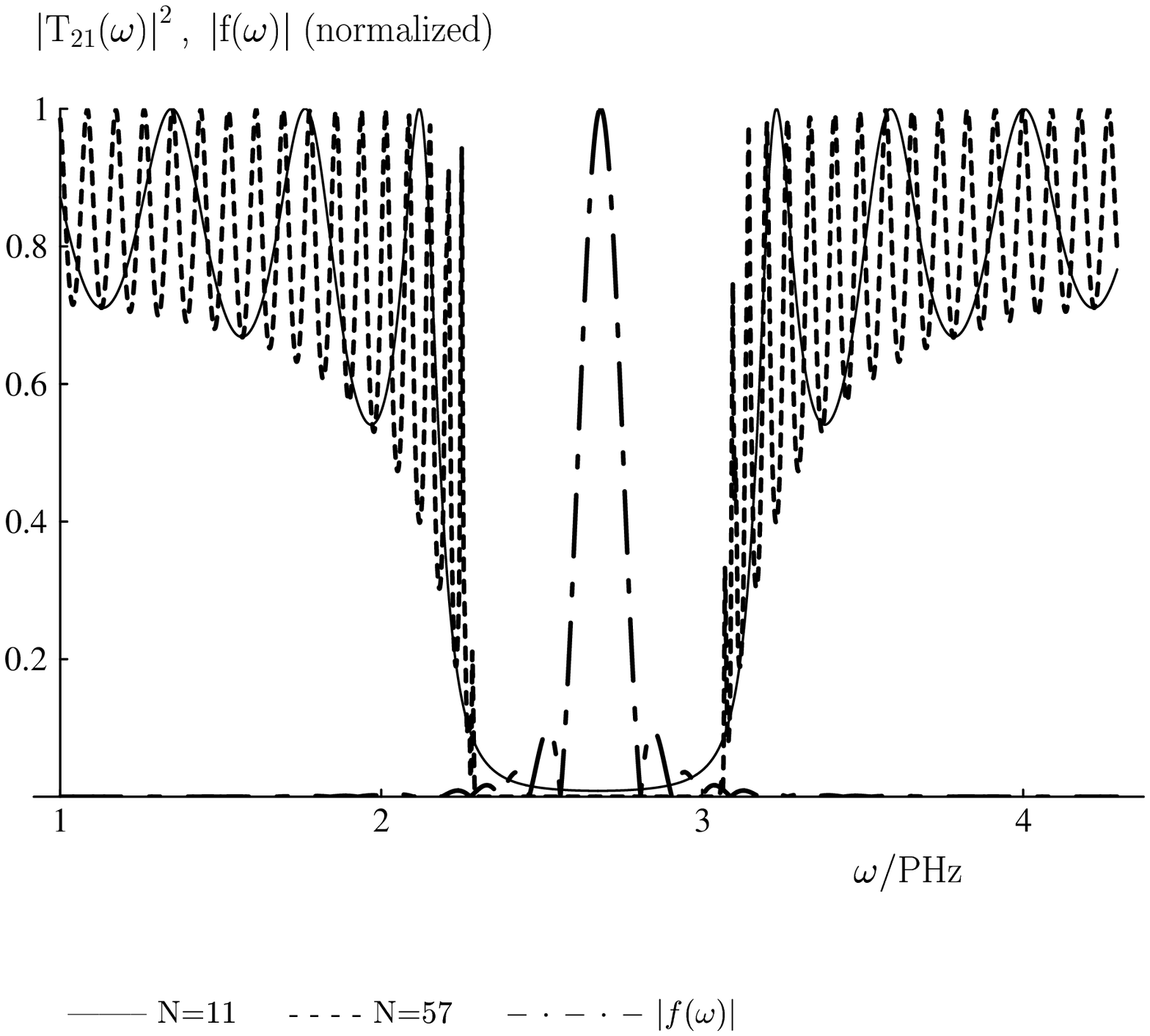,height=9in,angle=0}}
\end{picture}
\\ \\ \\ \\
\noindent
Figure 3: The square of the absolute value of the transmittance of a
multilayer non-absorbing barrier ($ n_{\rm TiO_2} $ $\!=$ $\! 2.22 $,
$ n_{\rm SiO_2} $ $\!=$ $\! 1.41 $), $|T_{21}(\omega)|^2$,
is shown for $N$ $\!=$ $\!11$ layers and $N$ $\!=$ $\!57$ layers
together with the (normalized) spectral line shape function of a pulse that is
assumed to be limited in time ($2t_0$ $\!=$ $\!40$\,fs). 
\label{fig3}

\newpage
\begin{picture}(70,500)
\put(-28,-100){\psfig{figure=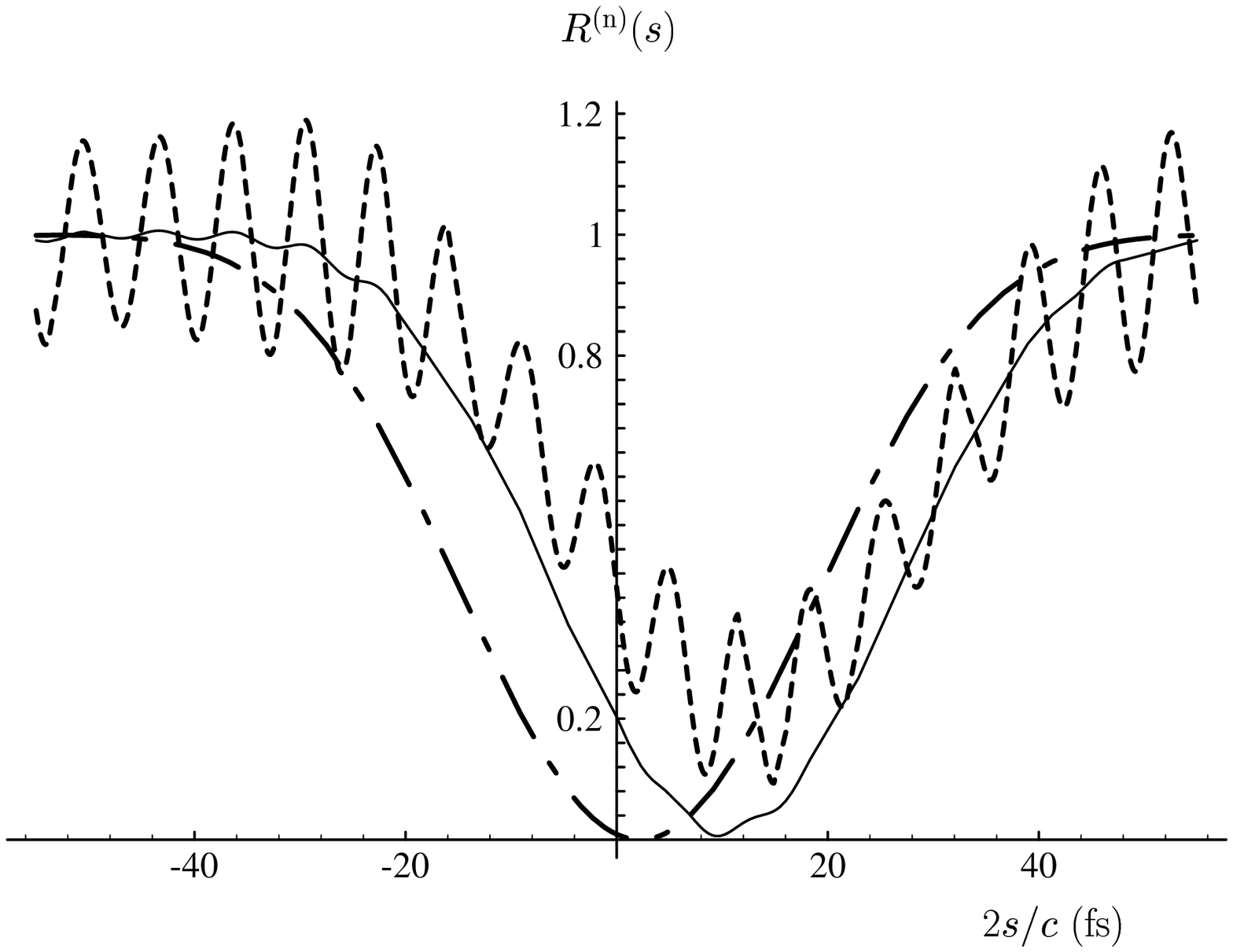,height=9in,angle=0}}
\end{picture}
\noindent
\\
Figure 4: The (normalized) coincidences $R^{({\rm n })}(s)$ 
are shown for an experimental scheme \protect\cite{CKS} 
in dependence on the
translation length $s$ for a time-limited pulse of the
incoming photon ($2t_0$ $\!=$ $\!40$\,fs) and
various numbers of the layers of a lossless barrier:
$N$ $\!=$ $\!11$ (dotted-dashed line),
$N$ $\!=$ $\!35$ (full line),
$N$ $\!=$ $\!41$ (dashed line).
The data of the lossless barrier are the same as in
Fig.~\protect3.

\newpage
\begin{picture}(70,500)
\put(-28,-100){\psfig{figure=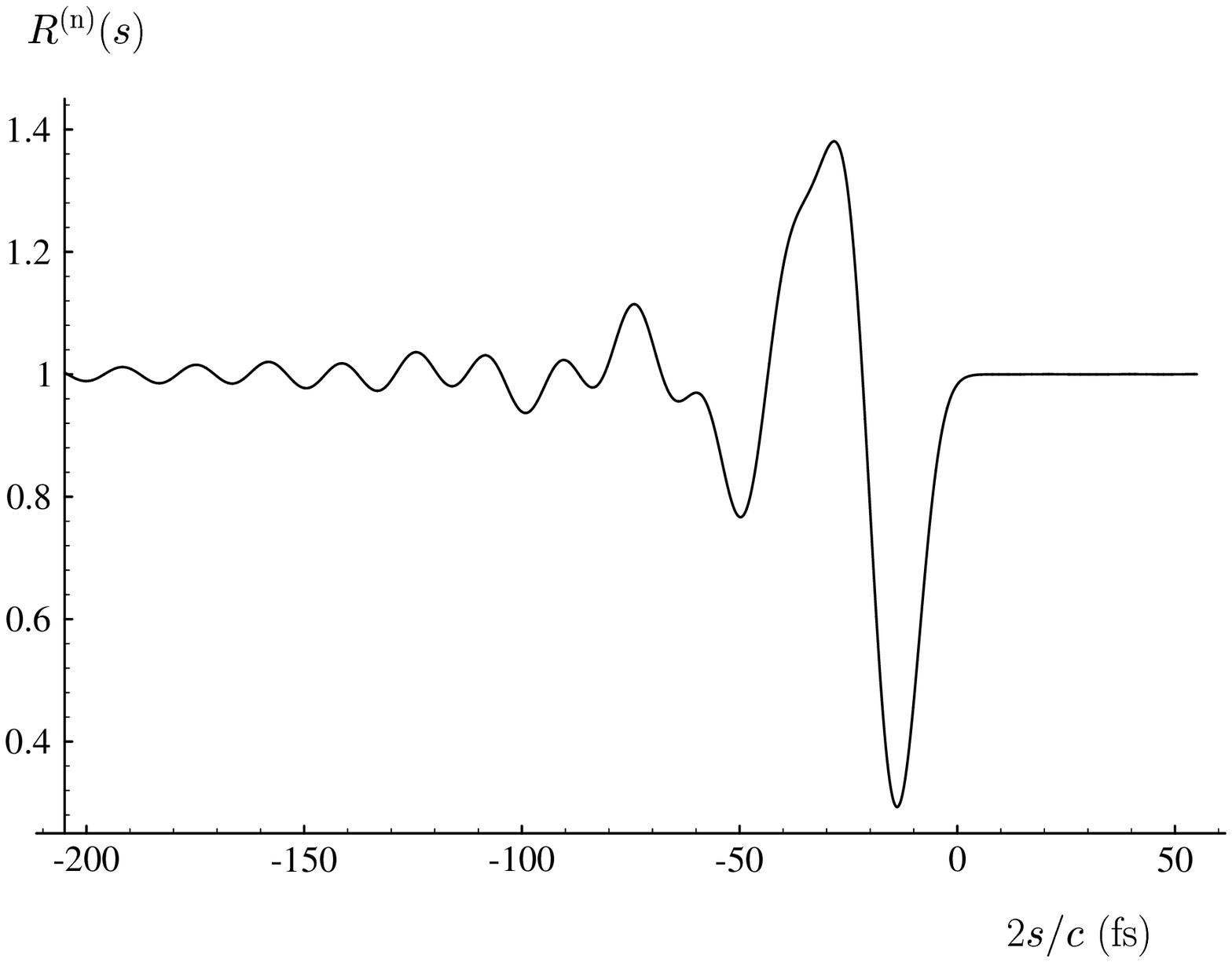,height=9in,angle=0}}
\end{picture}
\noindent
\\
Figure 5: The (normalized) coincidences $R^{({\rm n })}(s)$ 
are shown for an experimental scheme similar to \protect\cite{CKS}
but for a laser frequency $ \Omega = 6.22 $ PHz, 
in dependence on the
translation length $s$ for a time-limited pulse of the
incoming photon ($2t_0$ $\!=$ $\!40$\,fs) and
a Bragg mirror with
$N$ $\!=$ $\!57$ layers.
The data of the lossless barrier are the same as in
Fig.~\protect3.

\newpage
\begin{picture}(70,500)
\put(-28,-100){\psfig{figure=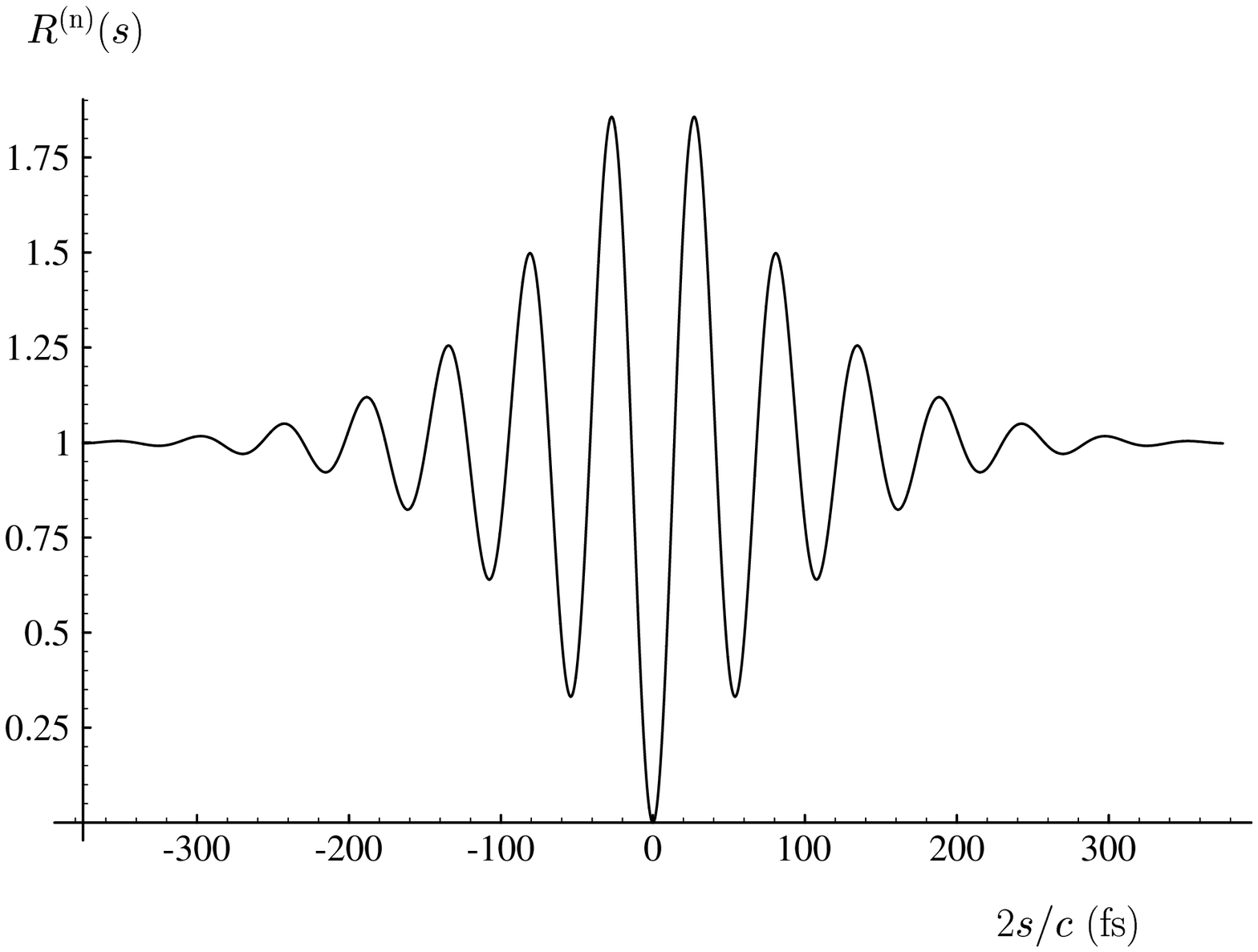,height=9in,angle=0}}
\end{picture}
\noindent
\\
Figure 6: The (normalized) coincidences $R^{({\rm n })}(s)$ 
in dependence on the
translation length $s$
are shown for the experimental scheme sketched in Fig.~\protect1 
including two identical
Bragg mirrors with
$N$ $\!=$ $\!57$ layers in path I and II.
A laser frequency tuned to $ \Omega = 6.16 $ PHz and 
time-limited pulses of the
incoming photons ($2t_0$ $\!=$ $\!40$\,fs) have been assumed.
The data of the lossless barriers are the same as in
Fig.~\protect3.

\end{document}